# Video shot boundary detection using motion activity descriptor

Abdelati Malek Amel,   Ben Abdelali Abdessalem and Mtibaa Abdellatif

**Abstract**—This paper focus on the study of the motion activity descriptor for shot boundary detection in video sequences. We interest in the validation of this descriptor in the aim of its real time implementation with reasonable high performances in shot boundary detection. The motion activity information is extracted in uncompressed domain based on adaptive rood pattern search (ARPS) algorithm. In this context, the motion activity descriptor was applied for different video sequence.

———————————— ◆ ————————————

## 1. INTRODUCTION

In recent years, the development of software and hardware technology has enabled the creation of a large amount of digital video content. Video segmentation based on motion [1] is a new research area. Motion is a salient feature in video, in addition to other typical image features such as color, shape and texture. Video shot boundary detection is a fundamental step in video indexing and retrieval, and in general video data management. The general objectives are to segment a given video sequence into its constituent shots, and to identify and classify the different shot transitions in the sequence [10]. Different algorithms have been proposed, for instance, based on simple color histograms [11, 12], pixel color differences [13], color ratio histograms [14], edges [15], and motion [16– 18]. In this work, we study the problem of video partitioning using a motion-based approach.

In this study we interest in the validation of the motion activity descriptor in the aim of its possible real time implementation. We have applied this descriptor for different video sequence. We have considered its application for shot boundary detection.

The rest of this paper is organized as follows: in section 2, we present the motion activity descriptor specification. In section 3 we focus on the study of the motion activity descriptor for shot boundary detection in video sequences. The different experimentations and the evolution results of the motion activity are presented in section 4.

## 2 MOTION ACTIVITY DESCRIPTOR

The motion activity is one of the motion features included in the visual part of the MPEG-7 standard. It also used to describe the level or intensity of activity, action, or motion in that video sequence. In this paper, we propose that since low or high action is a measure of how much a video segment is changing.

The motion activity is used in different applications such as video surveillance, fast browsing, dynamic video summarization, content-based querying etc. This information can for example be used for shot boundary detection, for shot classification or scene segmentation [19].

The magnitude of the motion vectors represents a measure of intensity of motion activity that includes

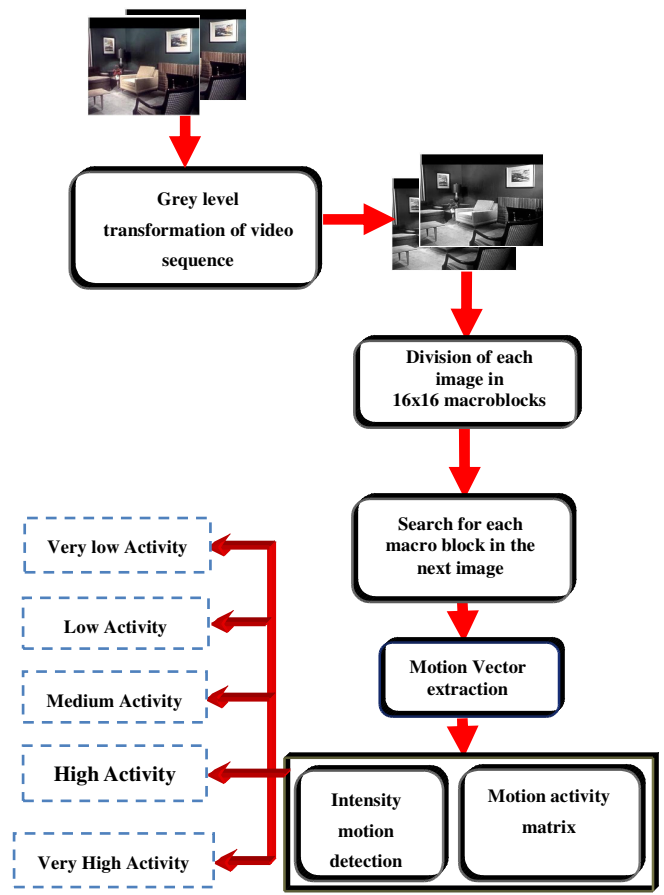

Fig.1. Different steps for motion intensity extraction

• *Abdelati Malek Amel is with the National school of engineers of monastir, Tunisia.*
• *Ben Abdelali Abdessalem is with the High Institute of Informatics and Mathematics of Monastir, Tunisia.*
• *Mtibaa Abdellatif is with the National School ef engineers, Monastir, Tunisia.*

several additional attributes that contribute towards the efficient use of these motion descriptors in a number of applications. Intensity of activity [4] is expressed by an integer in the range (1-5) and higher the value of intensity, higher the motion activity.

The motion features are extracted using the motion vectors. Block motion techniques are employed to extract the motion vectors. Suppose x (i, j) and y (i, j) denote the motion vectors in x and y directions for a given frame, where (i, j) indicates the block indices. The spatial activity matrix was defined in [5] to be

$$C_{mv} = \{R(i, j)\} \qquad (1)$$

Where

$$R_{xy}(i, j) = \sqrt{\left(x(i, j)^2 + y(i, j)^2\right)} \qquad (2)$$

The average of activity matrix for each frame is given by

$$C_{mv}^{avg} = \frac{1}{MN} \sum_{i=0}^{M-1} \sum_{j=0}^{N-1} C_{mv}(i, j) \qquad (3)$$

Here M and N are the width and height of the macro-blocks in the frame. This method ignores the low activity blocks and leaves the high activity blocks unaltered. The intensity of motion for each frame is determined as follows:

$$\sigma_{fr}^2 = \frac{1}{MN} \sum_{i=0}^{M-1} \sum_{j=0}^{N-1} \left(C_{mv}(i, j) - C_{mv}^{avg}\right)^2 \qquad (4)$$

Intensity values of motion are classified into five categories (Table 1) namely very low, low, medium, high and very high activity.

The intensity of motion—the level of motion activity and

#### TABLE 1
##### MOTION ACTIVITY CATEGORIES

| Activity value | Dynamique de l'écart-type des vecteurs de mouvement σ |
|---|---|
| 1/Very low Intensity | $0 \leq \sigma \leq 3,9$ |
| 2/Low Intensity | $3,9 \leq \sigma \leq 10,7$ |
| 3/Medium Intensity | $10,7 \leq \sigma \leq 17,1$ |
| 4/High Intensity | $17,1 \leq \sigma \leq 32$ |
| 5/Very High Intensity | $32 \leq \sigma$ |

it's their feature used for motion activity description. In the next section we describe how to extract motion intensity in the uncompressed domain. We use the block matching algorithm to extract the motion intensity feature.

### 2.1. Block_matching for motion vector extraction

To extract the motion intensity, the motion of the moving objects has to be estimated first. This is called motion estimation. The commonly used motion estimation technique in all the standard video codecs is the block matching algorithm (BMA). Block matching is used to retrieve an initial estimate of the image displacement. To obtain a dense displacement field, matching with adaptive block sizes was implemented. In this typical algorithm, a frame is divided into blocks of M × N pixels or, more usually, square blocks of N2 pixels [3]. Then, we assume that each block undergoes translation only with no scaling or rotation. The blocks in the first frame are compared to the blocks in the second frame. Motion Vectors can then be calculated for each block to see where each block from the first frame ends up in the second frame. A vast number of BMAs have been proposed:

- Exhaustive search (ES)
- Diamond Search(DS)
- Three-step search(TSS)
- New three step search(NTSS)
- Simple and effective search(SES)
- Four Step Search (4SS)
- Adaptive Rood Pattern Search(ARPS)

In this work, we use an adaptive rood pattern search (ARPS) algorithm.

A. Barjatya[7] illustrates and simulates 7 of the most popular block matching algorithms, with their comparative study. Three more, very recent, block matching algorithms are studied in the end as part of literature review. Of the various algorithms studied or simulated in [7], ARPS turns out to be the best block matching algorithm.

ARPS [8] algorithm makes use of the fact that the general motion in a frame is usually coherent, i.e. if the macro blocks around the current macro blocks moved in a particular direction then there is a high probability that the current macro block will also have a similar motion vector. This algorithm uses the motion vector of the macro block to its immediate left to predict its own motion vector. An example is shown in Fig. 1. The predicted motion vector points to (3, -2). In addition to checking the location pointed by the predicted motion vector, it also checks at a rood pattern distributed points, as shown in Fig 1, where they are at a step size of S = Max ( |X|, |Y| ). X and Y are the x-coordinate and y-coordinate of the predicted motion vector. This rood pattern search is always the first step.

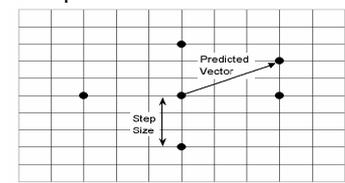

Fig. 2. Adaptive Root Pattern: The predicted motion vector is (3,-2) and the step size S = Max( |3|, |-2|) = 3.

It directly puts the search in an area where there is a high probability of finding a good matching block.

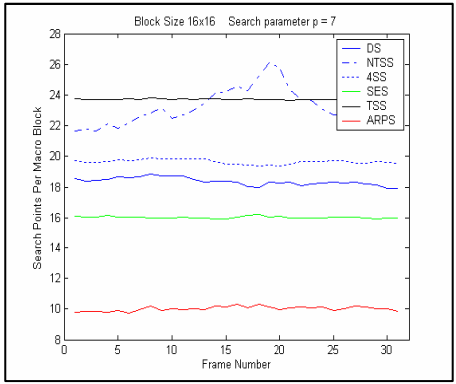

Fig. 3. Search points per macro block while computing the PSNR performance of Fast Block Matching Algorithms [7].

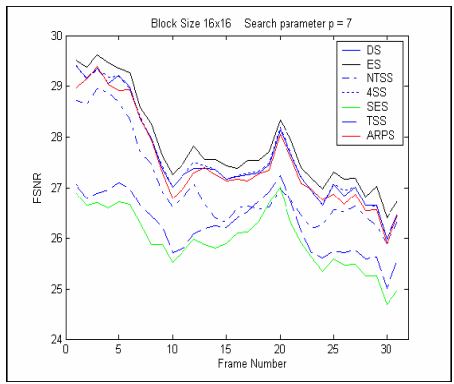

Fig. 4. Peak-Signal-to-Noise-Ratio (PSNR) performance of Fast Block Matching Algorithms. Caltrain Sequence was used with a frame distance of 2 [7].

Peak-Signal-to-Noise-Ratio (PSNR) characterizes the motion compensated image that is created by using motion vectors and macro clocks from the reference frame.

An example of the Matlab implementation results of the motion vector description technique is given by fig 5.

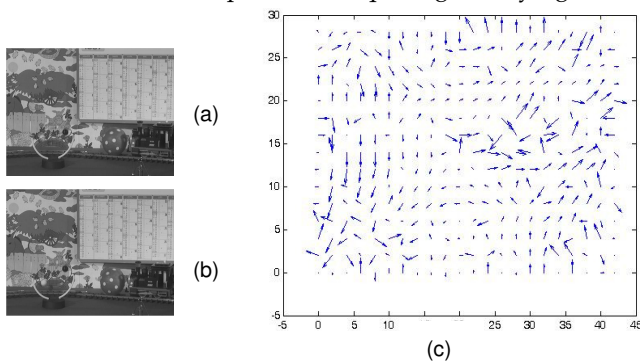

Fig.5. (a) and (b) two successive caltrain frames and (c) motion vector result of (a) and (b) frames.

### 2.2. Motion activity implementation for an example of video sequence

An example of spatial repartition of the motion activity descriptor is given in fig 6. The obtained results demonstrate the performance of the motion descriptor for "News-4" sequence.

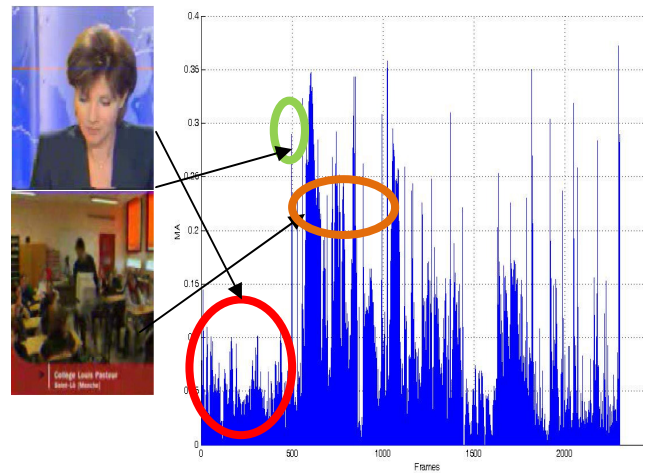

Fig.6. Example of video sequence results with the motion activity descriptor

## 3 SHOT BOUNDARY DETECTION USING MOTION ACTIVITY

The basis of any video segmentation method consists in detecting visual discontinuities along the time domain. During this process, it is required to extract visual features that measure the degree of similarity between frames in a given shot. This measure is related to the difference or discontinuity between frame n and n+j where j>= 1.

The main idea underlying the methods of segmentation schemes is that images in the vicinity of a transition are highly dissimilar. It then seeks to identify discontinuities in the video stream. The general principle is to extract a comment on each image, and then define a distance [9] (or similarity measure) between observations. The application of the distance between two successive images, the entire video stream, roduces a one-dimensional signal, in which we seek then the peaks (resp. hollow if similarity measure), which correspond to moments of high dissimilarity.

In this work, we used the extraction of key frames method based on detecting a significant change in the activity of motion. To jump 2 images which do not distort the calculations but we can minimize the execution time. First we extract the motion vectors between image i and image i+2 then calculates the intensity of motion, we repeat this process until reaching the last frame of the video and comparing the difference between the intensities of successive motion to a specified threshold. The idea can be visualized in fig 7.

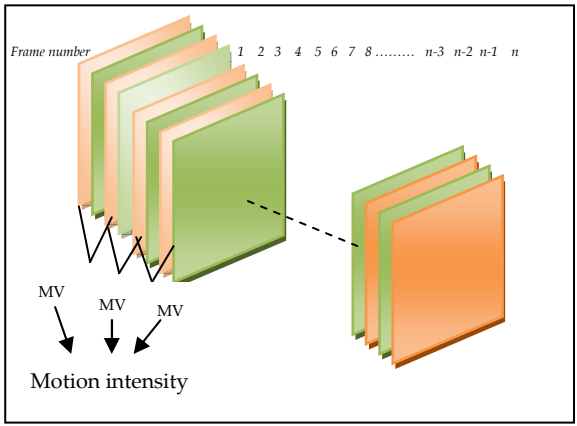

Fig.7.The idea of video segmentation using motion intensity

## 4 EXPERIMENTATION

The goal of short boundary detection is to process video sequences that contain high redundancy and make them exciting, interesting, valuable and useful for users.

Many researchers do not include any form of quantitative evaluations. The criteria to judge shot boundary detection quality will be different for different application domains. For event-based content, summaries might be scored using the percentage of events from the original program that they contain. Two evaluations are given in [8] and [9]. Ekin and Tekalp [9] give precision and recall values for goal, referee, and penalty box detection, which are important events in soccer games.

$$\text{Precision} = \text{Transitions Correctly Reported/Transitions Reported} \quad (5)$$

$$\text{Recall} = \text{Transitions Correctly Reported/Transitions in Reference} \quad (6)$$

$$\text{FP} = \text{False Transitions / Transitions Reported} \quad (7)$$

For every video sequence we determine the number of shots, the number of shots correctly reported, the number of false detections and the number of non reported shots. For each sequence we also draw the curve of the distances between the successive frames. These curves are mainly used to determine the threshold values, but they also give an idea about the capacity of the used technique in detecting transitions.

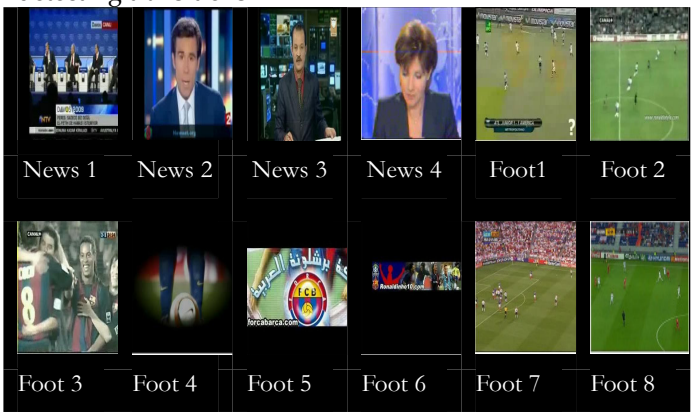

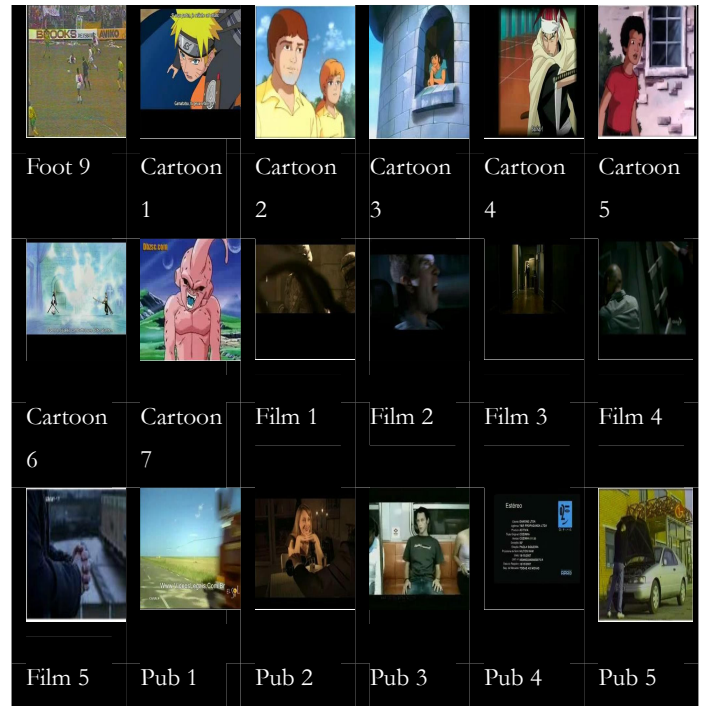

Fig.8. Example frames from the video sequence employed in the experimental evaluation.

An overview of our shot boundary detection results is given in Table2. The segmentation method does not allow to explicitly selecting the total number of homogeneous segments. This number depends on the threshold value adopted in the merging decision. We have studied the sensibility of our motion-based segmentation method to the threshold setting. The video segmentation using motion activity is given in Table2. With motion activity descriptor we obtain good overall results for shot boundary detection, as it is shown in Table3.

TABLE 2
MOTION VIDEO SEQUENCE PROPERTIES

| Type de vidéos | Durée (min) | Nombre d'images | Transitions totales | CUTs | Fondus |
|---|---|---|---|---|---|
| News | 05:56 | 7136 | 83 | 62 | 21 |
| Cartoon | 04:56 | 8255 | 110 | 89 | 21 |
| Football | 04:14 | 11220 | 134 | 99 | 35 |
| Films | 03:45 | 7733 | 60 | 57 | 3 |
| Publicity | 02:38 | 5962 | 90 | 60 | 30 |
| TOTAL | 21:29 | 40306 | 477 | 367 | 110 |

The obtained results for the different evaluation tests are given in Table 3. This table gives the performances (precision, recall and false positives) of the motion activity descriptor for different video sequence types.

TABLE 3
PERFORMANCES OF MOTION ACTIVITY DESCRIPTOR FOR
DIFFERENT VIDEO SEQUENCE

| Video | Precision | Recall | FP |
|---|---|---|---|
| News | 72,72% | 75,33% | 28,28% |
| Football | 69,56% | 84,21% | 30,43% |
| Films | 75% | 66,66% | 25% |
| Publicité | 76,19% | 84,21 % | 23,8% |
| Cartoons | 63,63% | 75,71% | 36,36% |

An example of graphical results (for the video "News 4") is given in Fig 9. The curve represented in this figure demonstrates the difference between two successive motion activity for 2 frames to obtain a shot boundary detection.

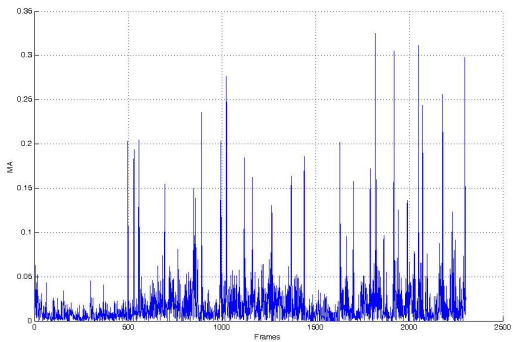

Fig.9. The difference between 2 successive motion intensities for shot boundary detection of the "news-4".

The first frame of each shot is considered as the key image of the detected shot. The Fig 10 represents an example of shot boundary detection corresponding to the sequence "News 4".

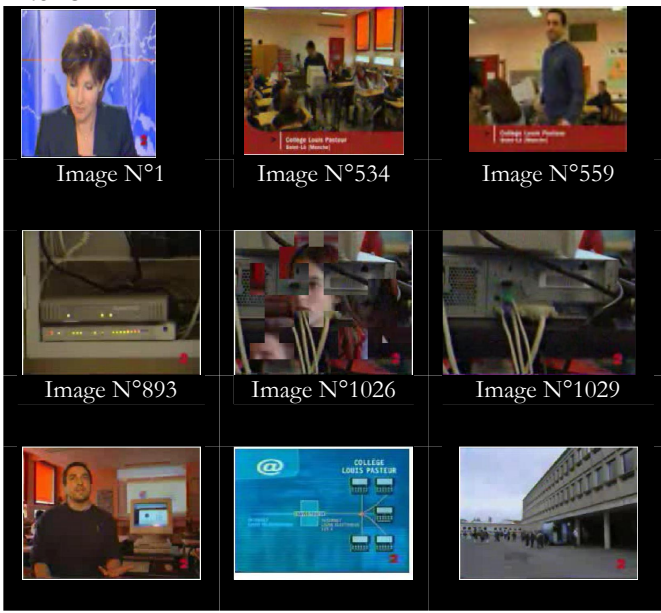

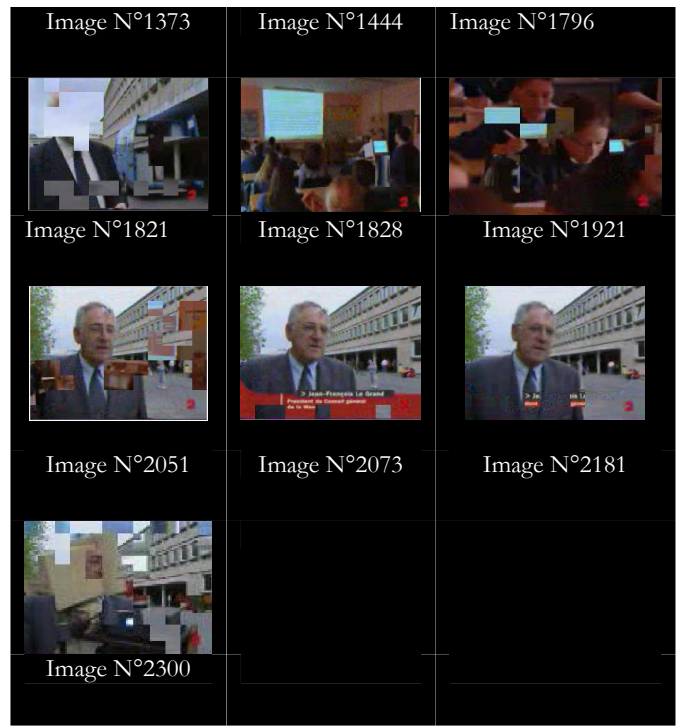

Fig.10. An example of shot boundary detection (the "News 4" video sequence)

## 5 CONCLUSIONS

This paper brought an experimental study of the motion activity descriptor particularly for shot boundary detection. After introducing the content based video segmentation problem, we have presented the specification of the motion activity in section 2. Section 3 and section 4 were dedicated to present the experiments of the motion activity descriptor for shot boundary detection in video sequences. The motion activity was applied for different video sequences. Experimental results demonstrate that the use of motion activity can assure a satisfactory of shot boundary detection rate. This can promise a better computing performance and can be useful for real time implementation.

### References

[1] Amel Malek Abdelati, Jlassi bahaeddine, Abdessalem Ben Abdellali and Abdellatif Mtibaa., " Suivi d'objet en movement par la method Level Set variationnelle, " international workshop on system engineering design & application, October 24-26, 2008, Monastir, Tunisia.

[2] ISO/IEC 15938-3 "Information Technology - Multimedia Content,"Description Interface, Part 3 Visual, 2001.

[3] ISHIGURO, T., and IINUMA, K. "Television bandwidth compression transmission by motion-compensated interframe coding," IEEE Commun. Mag., 1982, 10, pp.24–30

[4] S. Jeannin and A. Divakaran, "MPEG-7 visual motion descriptors, " IEEE Transactions on Circuits and Systems for Video Technology, vol. 11, pp. 720-724, 2001.

[5] D. A. X. Sun and B. S. Manjunath, "A Motion Activity Descriptor and Its Extraction in Compressed Domain," IEEE

*Pacific-Rim Conference on Multimedia (PCM)*, vol. LNCS 2195, pp. 450-453, October 2001.

[6] Yao Nie, and Kai-Kuang Ma, "Adaptive Rood Pattern Search for Fast Block-Matching Motion Estimation, " *IEEE Trans. Image Processing*, vol 11, no. 12, pp. 1442-1448,December 2002.

[7] Aroh Barjatya, "Block Matching Algorithms For Motion Estimation, "*Student Member, IEEE, DIP 6620 Spring 2004* Final Project Paper

[8] A.M. Ferman and A.M. Tekalp, "Two-stage hierarchical video summary extraction to match low-level user browsing preferences," *IEEE transactions on multimedia*, 5(25):244-256, 2003.

[9] Ahmet Ekin and A.Murat Tekalp, "Automatic Soccer video analysis and summarization, "*In Proceedings of SPIE Conference on storage and Retrieval for Media Databases 2003*, volume 5021, pages 339-350, Santa Clara, CA, 20-24 January 2003.

[10] Don Adjeroh, M. C. Lee, N. Banda, and Uma Kandaswamy, "Adaptive Edge-Oriented Shot Boundary Detection,*" Hindawi Publishing Corporation EURASIP Journal on Image and Video Processing* Volume 2009, Article ID 859371, 13 pages doi:10.1155/2009/859371, 18 May 2009.

[11] M. J. Swain and D. H. Ballard, "Color indexing,*" International Journal of Computer Vision*, vol. 7, no. 1, pp. 11–32, 1991.

[12] A. Nagasaka and Y. Tanaka, "Automatic video indexing and full-video search for object appearances," *in Visual Database System*s II, Elsevier, 1992.

[13] H. Zhang, A. Kankanhalli, and S. W. Smoliar, "Automatic partitioning of full-motion video," *Multimedia Systems*, vol. 1,no. 1, pp. 10–28, 1993.

[14] D. A. Adjeroh and M. C. Lee, "Robust and efficient transform domain video sequence analysis: an approach from the generalized color ratio model," *Journal of Visual Communication and Image Representation*, vol. 8, no. 2, pp. 182–207, 1997.

[15] R. Zabih, J. Miller, and K. Mai, "A feature-based algorithm for detecting and classifying production effects," *Multimedia Systems*, vol. 7, no. 2, pp. 119–128, 1999.

[16] J. D. Courtney, "Automatic video indexing via object motion analysis," *Pattern Recognition*, vol. 30, no. 4, pp. 607–625, 1997.

[17] P. Bouthemy, M. Gelgon, and F. Ganansia, "A unified approach to shot change detection and camera motion characterization," *IEEE Transactions on Circuits and Systems for Video Technology*, vol. 9, no. 7, pp. 1030–1044, 1999.

[18] S. Dagtas, W. Al-Khatib, A. Ghafoor, and R. L. Kashyap, "Models for motion-based video indexing and retrieval," *IEEE Transactions on Image Processing*, vol. 9, no. 1, pp. 88–101, 2000.

[19] Report ,"State of the Art of Content Analysis Tools for Video, Audio and Speech, " Werner Bailer (JRS), Franz Höller (JRS), Alberto Messina (RAI), Daniele Airola (RAI), Peter Schallauer (JRS), Michael Hausenblas (JRS), 10/3/2005